\documentclass[a4paper,aps,prd,10pt,preprintnumbers,showpacs,twocolumn,superscriptaddress,nofootinbib,amsmath,amssymb]{revtex4-1}

\usepackage{orcidlink,graphicx,cmap}
\usepackage{cmap}
\usepackage[utf8]{inputenc}
\usepackage[T1]{fontenc}

\begin{document}
\title{Particle Motion in Regular Black Hole Spacetimes Supported by a Galactic Halo}
\author{Bekir Can Lütfüoğlu}
\email{bekir.lutfuoglu@uhk.cz}
\affiliation{Department of Physics, Faculty of Science, University of Hradec Králové, Rokitanského 62/26, 500 03 Hradec Králové, Czech Republic}

\begin{abstract}
We investigate particle motion in regular and asymptotically flat black hole spacetimes supported by Dehnen-type dark-matter halos. Two analytic models are analyzed, allowing a systematic study of circular geodesics, photon-sphere properties, shadow radius, Lyapunov exponent, ISCO frequency, binding energy, and Hawking temperature. The corrected numerical results show that the halo scale parameter can significantly modify strong-field observables. In both models, for moderate density slopes, increasing the halo parameter reduces characteristic radii while enhancing orbital instability and accretion efficiency. For steeper density falloff, however, deviations from the Schwarzschild case remain small. These results demonstrate that halo-induced modifications of optical and dynamical black hole signatures are strongly controlled by the density profile parameters.
\end{abstract}

\maketitle

\section{Introduction}

Astrophysical black holes are not isolated systems but are embedded in galactic environments whose gravitational dynamics are dominated at large scales by dark matter. Observations ranging from galactic rotation curves to gravitational lensing and cosmological structure formation provide compelling evidence for extended dark-matter halos surrounding galaxies. While the microscopic nature of dark matter remains unknown, its macroscopic gravitational influence can be modeled effectively through phenomenological density profiles. Incorporating such halo distributions into black hole spacetimes provides a natural framework for studying environmental effects on strong-field gravity \cite{Mo:2010bk}.

In recent years, considerable attention has been devoted to constructing black hole solutions supported by realistic dark-matter profiles. In particular, it has been shown that certain halo models, when supplemented by suitable pressure prescriptions, can give rise to regular black hole geometries that avoid central curvature singularities. These configurations offer a theoretically appealing alternative to classical singular solutions and provide a laboratory for testing how the presence of halo matter modifies the near-horizon and strong-gravity regimes.

One of the most direct ways to probe the geometry of a compact object is through the analysis of particle motion in its vicinity. Circular geodesics and their stability properties encode essential information about the spacetime structure and determine key astrophysical observables. The innermost stable circular orbit (ISCO) sets the inner edge of thin accretion disks and directly influences the emitted radiation spectrum. The binding energy at the ISCO determines the efficiency of accretion processes, while the orbital frequency at the ISCO plays a central role in modeling high-frequency quasi-periodic oscillations. In addition, the Lyapunov exponent associated with unstable circular null geodesics quantifies the instability timescale of photon orbits and is closely related to the properties of black hole shadows and, in the eikonal limit, to quasinormal modes. The radius of the shadow itself provides a geometric observable that can be compared with direct imaging data from the Event Horizon Telescope and future interferometric missions.

In this work, we investigate these characteristic quantities for a regular black hole supported by a dark-matter halo. Our goal is to quantify how the halo parameter and the regularity of the interior modify the geodesic structure relative to the Schwarzschild case and to determine whether these modifications could produce observable signatures. By analyzing the Lyapunov exponent of unstable circular orbits, the ISCO radius and frequency, the binding energy, and the shadow radius within a unified framework, we aim to provide a comprehensive description of particle dynamics in this class of spacetimes.

The recent literature on particle motion, accretion, perturbations and optical phenomena in the background of black holes and other compact objects immersed in galactic halo is extensive 
 \cite{Kazempour:2024lcx,Liu:2023oab,Heydari-Fard:2024wgu,Tan:2024hzw,Qi:2026zrr,Heydari-Fard:2026hro,Nieto:2025apz,Pathrikar:2025sin,Dubinsky:2025fwv,Konoplya:2021ube,Lutfuoglu:2025kqp,Bolokhov:2025zva,Yue:2026evf,Yue:2026ish} emphasizing the importance of black holes in astrophysical environment for precise tests of gravity. 
 
Here we will study particle motion around regular black holes immersed in galactic halo suggested recently in \cite{Konoplya:2025ect} and further studied in \cite{Bolokhov:2025zva,Lutfuoglu:2025mqa}. The key finding in this context was that the black hole, otherwise singular, becomes regular because of the galactic environment \cite{Konoplya:2025ect}.

The paper is organized as follows. In Sec. \ref{sec:wavelike} we introduce the black hole spacetimes, while in Sec. \ref{sec:particlemotion} the particle motion is discussed. Finally, in Sec. \ref{Conclusion}, we summarize the obtained results.

\section{black hole background}
\label{sec:wavelike}

In this work we consider a class of asymptotically flat black hole solutions sourced by dark matter halos within the framework of Einstein gravity coupled to an anisotropic fluid. Such geometries were recently constructed in~\cite{Konoplya:2025ect}, where it was shown that realistic halo density profiles, when supplemented by the condition $P_r=-\rho$, can generate exact analytic spacetimes that are free of curvature singularities. These solutions provide a physically motivated alternative to ad hoc regular metrics, since the matter content is directly tied to empirically used galactic density distributions.

Among the various configurations presented in~\cite{Konoplya:2025ect}, we focus on two representative cases. The choice is motivated primarily by their analytic simplicity and transparent physical interpretation. Both admit closed-form expressions for the metric function, allow straightforward analysis of geodesics and perturbations, and retain a clear separation between the black hole mass scale and the halo scale parameter. At the same time, they illustrate two qualitatively distinct realizations of regular black hole geometries supported by Dehnen-type matter distributions,
\begin{equation}
\rho(r) = \rho_{0} \left(\frac{r}{a}\right)^{-\alpha} \left(1+\frac{r^{k}}{a^{k}}\right)^{-(\gamma-\alpha)/k}, \label{Dehnendensity}
\end{equation}
which is widely employed in modeling galactic halos \cite{Dehnen:1993uh,Taylor:2002zd}. The parameter $a$ determines the scale at which the density transitions from its central behavior to the outer power-law decay, while $\gamma$ controls the asymptotic falloff.

We adopt the standard static and spherically symmetric line element
\begin{equation}
ds^{2} = - f(r)\, dt^{2} + f^{-1}(r)\, dr^{2} + r^{2}\left(d\theta^{2}+\sin^{2}\theta\, d\phi^{2}\right), \label{metric}
\end{equation}
where the lapse function $f(r)$ encodes both the central compact object and the surrounding halo.

\subsection*{Model I}

The first configuration corresponds to the Dehnen profile with parameters $\gamma=4$, $\alpha=0$, and $k=1$. In this case, the metric function takes the particularly compact form \cite{Konoplya:2025ect}
\begin{equation}
f_{1}(r) = 1 - 2M\,\frac{r^{2}}{(r+a)^{3}}, \label{f1}
\end{equation}
where $M$ denotes the ADM mass and $a>0$ determines the characteristic size of the halo. 

This model is especially convenient because it smoothly interpolates between two simple limits. At large distances, expanding $(r+a)^{-3}$ shows that
\[
f_{1}(r) = 1 - \frac{2M}{r} + \mathcal{O}(r^{-2}),
\]
so the spacetime approaches the Schwarzschild solution asymptotically. In contrast, near the center one finds
\[
f_{1}(r)
=
1 - \frac{2M}{a^{3}}\, r^{2}
+ \mathcal{O}(r^{3}),
\]
which corresponds to an effective de Sitter core and guarantees regularity at $r=0$. Owing to its algebraic simplicity and clear limiting behavior, this configuration serves as a minimal regular black hole model supported by a galactic halo.

\subsection*{Model II}

The second example is constructed from a Dehnen-type density profile with $\alpha=0$ and $k=3$. In this case, the lapse function can be written as \cite{Konoplya:2025ect}
\begin{equation}
f_{2}(r) = 1 - \frac{2M}{r} \left[1-\left(\frac{r^3}{a^3}+1\right)^{1-\gamma/3}\right]. \label{f2}
\end{equation}

Although slightly more involved algebraically, this model preserves analytic tractability and exhibits qualitatively similar asymptotic behavior. For large $r$, the halo contribution decays sufficiently rapidly to ensure the standard Schwarzschild falloff, while near the origin the density remains finite, leading again to a nonsingular interior. Compared to the first model, the $k=3$ choice produces a steeper transition between the central region and the asymptotic regime, allowing us to test the sensitivity of physical observables to the detailed structure of the halo profile.

\medskip 

The two models selected here thus represent analytically simple yet physically meaningful realizations of regular black holes embedded in dark-matter halos, providing a convenient framework for studying particle dynamics and other observable characteristics.

We work in geometrized units $G=c=1$ and set $M=1$, thereby measuring all quantities in units of the black hole mass. This normalization allows a direct assessment of how the halo scale $a$ influences the geometry relative to the central compact object.

\section{Particle dynamics and characteristic orbits}\label{sec:particlemotion}

\subsection{General setup}

To analyze geodesic motion, we introduce the function  \cite{Konoplya:2020hyk,Lutfuoglu:2025ldc},
\begin{equation}
\mathcal{P}(r) \equiv \frac{f(r)}{r^2},
\end{equation}
which proves particularly convenient for describing circular null trajectories and for identifying the region where strong-field effects are most pronounced. 

The function $\mathcal{P}(r)$ encodes the competition between gravitational redshift (through $f(r)$) and centrifugal effects ($1/r^2$). Its extrema determine special photon orbits and therefore delimit the region where radiation processes associated with unstable circular motion are expected to dominate. In this sense, we shall refer to the vicinity of such extrema as the \emph{radiation zone}, excluding both the near-horizon layer and the asymptotic weak-field region.

\begin{table}
\begin{tabular}{l c c c c c c c}
\hline
$a$ & $r_0$ & $T_H$ & $r_{m}$ & $R_{s}$ & $\lambda$ & $\Omega_{ISCO}$ & BE\\
\hline
$0$ & $2.000$ & $0.03979$ & $3.000$ & $5.19615$ & $0.192450$ & $0.06804$ & $0.05719$\\
$0.02$ & $1.939$ & $0.03978$ & $2.919$ & $5.09116$ & $0.193727$ & $0.07052$ & $0.05882$\\
$0.04$ & $1.877$ & $0.03973$ & $2.837$ & $4.98389$ & $0.194987$ & $0.07319$ & $0.06057$\\
$0.06$ & $1.814$ & $0.03966$ & $2.752$ & $4.87412$ & $0.196215$ & $0.07610$ & $0.06245$\\
$0.08$ & $1.749$ & $0.03953$ & $2.665$ & $4.76157$ & $0.197396$ & $0.07927$ & $0.06450$\\
$0.1$ & $1.682$ & $0.03935$ & $2.576$ & $4.64593$ & $0.198505$ & $0.08275$ & $0.06672$\\
$0.12$ & $1.613$ & $0.03909$ & $2.484$ & $4.52679$ & $0.199511$ & $0.08659$ & $0.06916$\\
$0.14$ & $1.541$ & $0.03874$ & $2.389$ & $4.40368$ & $0.200368$ & $0.09087$ & $0.07186$\\
$0.16$ & $1.466$ & $0.03826$ & $2.290$ & $4.27598$ & $0.201008$ & $0.09567$ & $0.07485$\\
$0.18$ & $1.387$ & $0.03760$ & $2.186$ & $4.14288$ & $0.201330$ & $0.10112$ & $0.07821$\\
$0.2$ & $1.303$ & $0.03669$ & $2.077$ & $4.00332$ & $0.201178$ & $0.10738$ & $0.08203$\\
$0.22$ & $1.213$ & $0.03539$ & $1.960$ & $3.85580$ & $0.200291$ & $0.11470$ & $0.08645$\\
$0.24$ & $1.114$ & $0.03345$ & $1.835$ & $3.69811$ & $0.198207$ & $0.12345$ & $0.09164$\\
$0.26$ & $1.00$ & $0.03031$ & $1.695$ & $3.52670$ & $0.194006$ & $0.13425$ & $0.09794$\\
$0.28$ & $0.855$ & $0.02418$ & $1.534$ & $3.33510$ & $0.185498$ & $0.14821$ & $0.10587$\\
\hline
\end{tabular}
\caption{Event horizon position ($r_0$), Hawking temperature ($T_H$), photon-sphere radius ($r_m$), shadow radius ($R_s$), Lyapunov exponent ($\lambda$), ISCO frequency ($\Omega_{ISCO}$) and binding energy (BE) for the first model ($M=1$).}
\end{table}

\begin{table}
\begin{tabular}{l c c c c c c c}
\hline
$a$ & $r_0$ & $T_H$ & $r_{m}$ & $R_{s}$ & $\lambda$ & $\Omega_{ISCO}$ & BE\\
\hline
$0$ & $2.000$ & $0.03979$ & $3.000$ & $5.19615$ & $0.192450$ & $0.06804$ & $0.05719$\\
$0.02$ & $1.789$ & $0.04186$ & $2.699$ & $4.75066$ & $0.204537$ & $0.07733$ & $0.06248$\\
$0.04$ & $1.693$ & $0.04275$ & $2.563$ & $4.55110$ & $0.210145$ & $0.08228$ & $0.06525$\\
$0.06$ & $1.614$ & $0.04341$ & $2.453$ & $4.39027$ & $0.214691$ & $0.08673$ & $0.06771$\\
$0.08$ & $1.545$ & $0.04392$ & $2.355$ & $4.24846$ & $0.218668$ & $0.09102$ & $0.07005$\\
$0.1$ & $1.480$ & $0.04432$ & $2.265$ & $4.11797$ & $0.222251$ & $0.09534$ & $0.07239$\\
$0.12$ & $1.418$ & $0.04461$ & $2.179$ & $3.99473$ & $0.225512$ & $0.09977$ & $0.07476$\\
$0.14$ & $1.358$ & $0.04476$ & $2.095$ & $3.87619$ & $0.228469$ & $0.10441$ & $0.07721$\\
$0.16$ & $1.298$ & $0.04476$ & $2.014$ & $3.76054$ & $0.231101$ & $0.10933$ & $0.07979$\\
$0.18$ & $1.238$ & $0.04456$ & $1.932$ & $3.64634$ & $0.233351$ & $0.11463$ & $0.08253$\\
$0.2$ & $1.176$ & $0.04408$ & $1.850$ & $3.53230$ & $0.235111$ & $0.12040$ & $0.08547$\\
\hline
\end{tabular}
\caption{Event horizon position ($r_0$), Hawking temperature ($T_H$), photon-sphere radius ($r_m$), shadow radius ($R_s$), Lyapunov exponent ($\lambda$), ISCO frequency ($\Omega_{ISCO}$) and binding energy (BE) for the second model at $\gamma=3.5$ ($M=1$).}
\end{table}

\begin{table}
\begin{tabular}{l c c c c c c c}
\hline
$a$ & $r_0$ & $T_H$ & $r_{m}$ & $R_{s}$ & $\lambda$ & $\Omega_{ISCO}$ & BE\\
\hline
$0$ & $2.000$ & $0.03979$ & $3.000$ & $5.19615$ & $0.192450$ & $0.06804$ & $0.05719$\\
$0.02$ & $1.980$ & $0.03978$ & $2.973$ & $5.16124$ & $0.192873$ & $0.06885$ & $0.05772$\\
$0.04$ & $1.959$ & $0.03977$ & $2.946$ & $5.12576$ & $0.193286$ & $0.06969$ & $0.05827$\\
$0.06$ & $1.938$ & $0.03975$ & $2.918$ & $5.08968$ & $0.193687$ & $0.07056$ & $0.05884$\\
$0.08$ & $1.917$ & $0.03971$ & $2.889$ & $5.05298$ & $0.194073$ & $0.07146$ & $0.05943$\\
$0.1$ & $1.894$ & $0.03967$ & $2.860$ & $5.01561$ & $0.194442$ & $0.07240$ & $0.06004$\\
$0.12$ & $1.872$ & $0.03960$ & $2.830$ & $4.97755$ & $0.194792$ & $0.07338$ & $0.06067$\\
$0.14$ & $1.849$ & $0.03952$ & $2.800$ & $4.93875$ & $0.195117$ & $0.07440$ & $0.06133$\\
$0.16$ & $1.825$ & $0.03942$ & $2.769$ & $4.89916$ & $0.195416$ & $0.07546$ & $0.06201$\\
$0.18$ & $1.800$ & $0.03930$ & $2.737$ & $4.85873$ & $0.195681$ & $0.07658$ & $0.06272$\\
$0.2$ & $1.775$ & $0.03916$ & $2.704$ & $4.81741$ & $0.195909$ & $0.07774$ & $0.06346$\\
\hline
\end{tabular}
\caption{Event horizon position ($r_0$), Hawking temperature ($T_H$), photon-sphere radius ($r_m$), shadow radius ($R_s$), Lyapunov exponent ($\lambda$), ISCO frequency ($\Omega_{ISCO}$) and binding energy (BE) for the second model at $\gamma =4$ ($M=1$).}
\end{table}

\begin{table}
\begin{tabular}{l c c c c c c c}
\hline
$a$ & $r_0$ & $T_H$ & $r_{m}$ & $R_{s}$ & $\lambda$ & $\Omega_{ISCO}$ & BE\\
\hline
$0$ & $2.000$ & $0.03979$ & $3.000$ & $5.19615$ & $0.192450$ & $0.06804$ & $0.05719$\\
$0.02$ & $2.000$ & $0.03978$ & $3.000$ & $5.19592$ & $0.192444$ & $0.06805$ & $0.05719$\\
$0.04$ & $1.999$ & $0.03977$ & $2.999$ & $5.19523$ & $0.192427$ & $0.06807$ & $0.05720$\\
$0.06$ & $1.998$ & $0.03975$ & $2.998$ & $5.19407$ & $0.192399$ & $0.06810$ & $0.05722$\\
$0.08$ & $1.997$ & $0.03972$ & $2.996$ & $5.19245$ & $0.192358$ & $0.06814$ & $0.05725$\\
$0.1$ & $1.995$ & $0.03969$ & $2.994$ & $5.19036$ & $0.192306$ & $0.06819$ & $0.05728$\\
$0.12$ & $1.993$ & $0.03964$ & $2.992$ & $5.18780$ & $0.192241$ & $0.06826$ & $0.05732$\\
$0.14$ & $1.990$ & $0.03959$ & $2.989$ & $5.18477$ & $0.192164$ & $0.06834$ & $0.05736$\\
$0.16$ & $1.987$ & $0.03953$ & $2.986$ & $5.18126$ & $0.192073$ & $0.06843$ & $0.05742$\\
$0.18$ & $1.984$ & $0.03945$ & $2.982$ & $5.17727$ & $0.191969$ & $0.06854$ & $0.05748$\\
$0.2$ & $1.980$ & $0.03937$ & $2.977$ & $5.17278$ & $0.191850$ & $0.06866$ & $0.05755$\\
\hline
\end{tabular}
\caption{Event horizon position ($r_0$), Hawking temperature ($T_H$), photon-sphere radius ($r_m$), shadow radius ($R_s$), Lyapunov exponent ($\lambda$), ISCO frequency ($\Omega_{ISCO}$) and binding energy (BE) for the second model at $\gamma=5$ ($M=1$).}
\end{table}

The four-momentum of a particle with rest mass $m$ is
\begin{equation}
p^\mu = m \frac{dx^\mu}{ds},
\end{equation}
where $s$ denotes the invariant affine parameter. The conserved quantities associated with stationarity and spherical symmetry are
\begin{equation}
E \equiv -p_t, \qquad L \equiv p_\phi .
\end{equation}

The normalization condition,
\begin{equation}
p_\mu p^\mu = -m^2,
\end{equation}
leads to the radial equation of motion. Restricting to the equatorial plane ($\theta=\pi/2$), the dynamics reduces to
\begin{equation}
m^2 \frac{1}{f(r)} \left(\frac{dr}{ds}\right)^2  = V_{\rm eff}(r),
\end{equation}
with the effective potential
\begin{equation}
V_{\rm eff}(r) = \frac{E^2}{f(r)} - \frac{L^2}{r^2} - m^2 .
\end{equation}

The radial motion is therefore governed by the sign and curvature of $V_{\rm eff}(r)$, which determine turning points, circular trajectories, and their stability.


\subsection{Circular trajectories}

Circular motion corresponds to stationary points of the effective potential,
\begin{equation}
V_{\rm eff}(r)=0, \qquad \frac{dV_{\rm eff}}{dr}=0 .
\end{equation}

Solving these conditions yields the conserved quantities associated with a circular orbit of radius $r$,
\begin{equation}
E^2 = -m^2 \frac{2 r \mathcal{P}^2(r)}{\mathcal{P}'(r)}, \qquad L^2 = -m^2 \frac{f'(r)}{\mathcal{P}'(r)} .
\end{equation}

An important invariant observable is the angular frequency measured at infinity,
\begin{equation}
\Omega^2 = \left(\frac{d\phi}{dt}\right)^2 = \frac{L^2 f^2(r)}{E^2 r^4} = \frac{f'(r)}{2r}.
\end{equation}

Remarkably, the orbital frequency depends only on the local gradient of the metric function and is independent of the particle mass.


\subsection{Null geodesics and the photon sphere}

In the massless limit, the structure simplifies considerably. The circular null orbit is determined solely by the extremum condition
\begin{equation}
\mathcal{P}'(r_m)=0.
\end{equation}

For physically relevant black hole spacetimes this extremum is a maximum, corresponding to the unstable photon sphere.

The impact parameter of photons,
\begin{equation}
b = \frac{L}{E},
\end{equation}
evaluated at the circular null orbit determines the shadow radius,
\begin{equation}
\frac{1}{R_s^2} = \frac{E^2}{L^2}
= \mathcal{P}(r_m).
\end{equation}

Thus, the shadow is directly controlled by the maximal value of $\mathcal{P}(r)$, demonstrating that both optical appearance and eikonal quasinormal modes are governed by the same geometric structure.


\subsection{Instability of the photon orbit}

To quantify the instability of the circular null trajectory, we consider small radial perturbations,
\begin{equation}
r(t) = r_m + \delta r(t).
\end{equation}

Expanding the radial equation to quadratic order gives
\begin{equation}
\left(\frac{d}{dt}\delta r\right)^2 = \lambda^2 \delta r^2 + \mathcal{O}(\delta r^3),
\end{equation}
where the Lyapunov exponent is
\begin{equation}
\lambda^2 = -\frac{1}{2 \mathcal{P}} \frac{d^2 \mathcal{P}}{dr_*^2} \Biggr|_{r=r_m}.
\end{equation}
Here $r_*$ denotes the tortoise coordinate, defined by
\begin{equation}
d r_{*} = \frac{d r}{f(r)}.
\end{equation}

The sign of $\lambda^2$ reflects the fact that the photon sphere corresponds to a maximum of $\mathcal{P}(r)$ and therefore represents an unstable equilibrium. This instability controls the imaginary part of eikonal quasinormal modes and sets the decay timescale of high-frequency perturbations.


\subsection{Stability of massive circular motion}

For massive particles, stability is governed by the second derivative of the effective potential. Expanding near a circular orbit,
\begin{equation}
m^2 \frac{1}{f(r)} 
\left(\frac{d\,\delta r}{ds}\right)^2
=
\frac{1}{2} V_{\rm eff}''(r)\,\delta r^2
+ \mathcal{O}(\delta r^3).
\end{equation}

The orbit is stable whenever
\begin{equation}
V_{\rm eff}''(r) < 0,
\end{equation}
which, after using the circularity conditions, can be written as
\begin{equation}
V_{\rm eff}''(r)
= -4 m^2 \frac{1}{r} 
\left(\frac{d \ln E}{dr}\right).
\end{equation}

The marginally stable orbit satisfies
\begin{equation}
V_{\rm eff}''(r_{ISCO}) = 0,
\end{equation}
defining the innermost stable circular orbit (ISCO) in asymptotically flat geometries.

At this radius, the particle energy attains its minimum,
\begin{equation}
E_{ISCO}^2 
= -m^2 
\max \left( \frac{2r \mathcal{P}^2(r)}{\mathcal{P}'(r)} \right).
\end{equation}

The binding energy,
\begin{equation}
BE = 1 - \frac{E_{ISCO}}{m},
\end{equation}
quantifies the efficiency of accretion and is one of the key observationally relevant parameters of the spacetime.


\subsection{Surface gravity and temperature}

Finally, the Hawking temperature is determined by the surface gravity at the event horizon $r_0$,
\begin{equation}
T_H = \frac{f'(r_0)}{4\pi}.
\end{equation}

This quantity sets the characteristic scale for quantum emission and provides a thermodynamic counterpart to the geometric properties governing particle motion.

\subsection{Interpretation of the Results}

The corrected numerical results presented in Tables~I--IV reveal a clear and internally consistent dependence of the characteristic quantities on the halo scale parameter $a$, with distinct behavior for the two models.

For Model~I (Table~I), increasing $a$ leads to a monotonic decrease of the event-horizon radius $r_0$, photon-sphere radius $r_m$, and shadow radius $R_s$. At the same time, both the Lyapunov exponent $\lambda$ and the ISCO frequency $\Omega_{ISCO}$ increase, and the binding energy grows noticeably. This indicates that in this configuration the halo effectively enhances the strength of the gravitational field in the near-horizon region: the photon sphere shifts inward, null circular orbits become more unstable (larger $\lambda$), and the efficiency of accretion increases (larger binding energy). The Hawking temperature decreases gradually and becomes significantly smaller for sufficiently large $a$, consistently reflecting the shrinking horizon size.

For the second model, the corrected data show that for $\gamma = 3.5$ (Table~II) the qualitative behavior is similar to Model~I: increasing $a$ reduces $r_0$, $r_m$, and $R_s$, while $\lambda$, $\Omega_{ISCO}$, and the binding energy increase. The Hawking temperature rises slightly at small $a$ and then remains nearly constant or decreases mildly at larger values. Thus, for moderate density slopes the halo again strengthens the effective gravitational field in the strong-field region.

For $\gamma = 4$ (Table~III), the same qualitative trend persists, although the magnitude of deviations from the Schwarzschild case is smaller. The characteristic radii decrease with $a$, while instability and accretion efficiency increase. In contrast, for $\gamma = 5$ (Table~IV), all deviations remain very small over the considered range of $a$: the horizon radius, photon-sphere position, shadow size, Lyapunov exponent, ISCO frequency, and binding energy change only marginally. This demonstrates that a steeper asymptotic density falloff strongly suppresses halo-induced modifications.

Overall, the corrected results indicate that in the second model the halo does not generically weaken the gravitational field. Instead, for moderate values of $\gamma$ it produces behavior qualitatively similar to Model~I, while sufficiently steep density profiles render the geometry effectively indistinguishable from the Schwarzschild case. Both the scale parameter $a$ and the slope parameter $\gamma$ therefore control the magnitude of deviations in optical and dynamical observables.

\section{Conclusions} \label{Conclusion}

In this work, we have investigated particle motion in regular black hole spacetimes supported by Dehnen-type dark-matter halos. Two analytic models were considered, allowing a systematic evaluation of the event-horizon radius, Hawking temperature, photon-sphere position, shadow size, Lyapunov exponent, ISCO frequency, and binding energy.

The numerical analysis shows that the halo scale parameter can substantially modify strong-field characteristics. In Model~I, increasing $a$ reduces the characteristic radii while enhancing orbital instability and accretion efficiency. In the second model, the corrected results demonstrate that for moderate density slopes ($\gamma = 3.5$ and $\gamma = 4$) the qualitative behavior is similar: characteristic radii decrease, while $\lambda$, $\Omega_{ISCO}$, and the binding energy increase. However, for steeper density profiles ($\gamma = 5$) the halo-induced corrections become very small, and the spacetime remains close to the Schwarzschild limit.

These findings indicate that environmental effects produced by realistic halo configurations can influence both optical signatures, such as the shadow size, and dynamical characteristics, such as ISCO properties and instability timescales. At the same time, the magnitude of these effects depends sensitively on the asymptotic density slope, which controls how efficiently the halo modifies the strong-field region.

\begin{acknowledgments}
BCL is grateful to Excellence Project PrF UHK 2205/2025-2026 for the financial support.
\end{acknowledgments}

\bibliography{bibliography}
\end{document}